# Fundamental Extended Objects for Chern-Simons Supergravity[1]

Pablo MORA[2] and Hitoshi NISHINO[3]

*Department of Physics
University of Maryland at College Park
College Park, MD 20742-4111, USA*

### Abstract

We propose a class of models in which extended objects are introduced in Chern-Simons supergravity in such a way that those objects appear on the same footing as the target space. This is motivated by the idea that branes are already first quantized object, so that it is desirable to have a formalism that treats branes and their target space in a similar fashion. Accordingly, our models describe interacting branes, as gauge systems for supergroups. We also consider the case in which those objects have boundaries, and discuss possible links to superstring theory and/or M-theory, by studying the fermionic $\kappa$-symmetry of the action.

---

[1]This work is supported in part by NSF grant # PHY-93-41926.
[2]Also at Instituto de Física, Facultad de Ciencias, Iguá 4225, Montevideo, Uruguay.
E-Mail: pmora@physics.umd.edu
[3]E-Mail: nishino@nscpmail.physics.umd.edu

## 1. Introduction

Gauge supergravity theories, based on lagrangians of the Chern-Simons form involving gauge fields for a gauge supergroup containing the super Poincaré group of the appropriate dimension, have recently attracted a great deal of attention. Since the construction of (2+1)-dimensional theories [1][2][3] and its quantization [2] these theories have been extended to higher dimensions by Chamseddine [4] and further developed by Bañados, Troncoso and Zanelli [5][6]. Also Green [7], or Moore and Seiberg [8] gave possible links between Chern-Simons theory for a supergroup and superstring theory, while Hořava [9] suggested that M-theory [10][11][12] may actually be a field theory of this class.

In this Letter we will suggest a natural way to introduce fundamental extended objects in Chern-Simons supergravities by embedding lower dimensional lagrangians for Chern-Simons supergravity in higher dimensions. We may also think of this model as a set of Chern-Simons supersymmetric branes[4] embedded into a background described also by a Chern-Simons action (and can be considered itself a Chern-Simons brane, even though it is not embedded in a larger manifold) providing in effect a 'brane-target space democracy'. It is important to remark, however, that supersymmetry is introduced here as a gauge symmetry, in the sense that the super Poincaré group (or any of it extensions, like the super conformal or super (anti) de Sitter groups) are realized as parts of a gauge group. That is to be contrasted with the standard approach of getting spacetime supersymmetry by defining the extended objects as embeddings in flat or curved superspace.

A heuristic motivation for the kind of models presented here is that if there is some underlying pregeometric theory in which the de Rham complex get physical content through some kind of antighost field variables, we must expect the effective action to be a sum of pieces of diverse dimensions of the form considered here. We believe that this simple observation is the key to understand the existence of extended objects in a fundamental theory of nature.

We also consider the case in which the embedded supersymmetric branes have boundaries. In such a case, we need to add boundary terms in order to preserve gauge invariance in a way resembling the action of Dixon *et al.* [14] for extended objects coupled to gauge fields except in two very important respects: First, we implement supersymmetry as a part of the gauge (super)group. Second, we have a bulk(s) with nontrivial degrees of freedom, which can be ignored only in the case that the gauge potential for the supergroup is pure-gauge. We will show that in this case and for a particular choice of the gauge supergroup one can establish a link with ten-dimensional (10D) superstring theory [15].

## 2. Action for Extended Objects

We consider a (super)group $G$ with generators $T^I$, the gauge potential 1-form $A = A_m{}^I T_I \, dx^m$ and the curvature 2-form $F = dA + A^2$, defined on a manifold $M^{D+1}$ of even

---

[4]Chern-Simons bosonic branes in a somewhat different framework have been considered in ref. [13].

dimension $D+1$. If we have an invariant polynomial $P(F)$ defined as the formal sum

$$P(F) = \sum_{n=1}^{(D+1)/2} \alpha_n \operatorname{STr}(F^n) \ , \tag{1}$$

here $\operatorname{STr}(T^{I_1} \ldots T^{I_n}) = g^{I_1 \ldots I_n}$ stands for an invariant symmetric trace on the algebra of $G$, we define the action for our system by

$$S = \sum_{d \text{ even}}^{D-1} \alpha_d \int_{\Omega^{d+1}} I_{d+1}^0 \ . \tag{2}$$

Here the $I_{d+1}^0(F,A)$ are the Chern-Simons forms associated with $P(F)$, so locally

$$P(F) = \sum_{d \text{ even}}^{D-1} \alpha_d \operatorname{STr}\left(F^{\frac{d+2}{2}}\right) = \sum_{d \text{ even}}^{D-1} \alpha_d \, dI_{d+1}^0(F,A) \ , \quad (n = \tfrac{d}{2}+1) \ , \tag{3}$$

and $M^{D+1}$ has $\Omega^D$ as its boundary, while the $\Omega^{d+1}$'s are submanifolds of $\Omega^D$ of odd dimension. The dynamical variables are the gauge potentials $A$ and the embedding coordinates of the submanifolds $X_{(d)}^m(\chi_{(d)}^i)$, where the $\chi_{(d)}^i$ with $i = 1, \ldots, d+1$ are local coordinates in $\Omega^{d+1}$. Notice that all these manifolds are supposed to be non compact at least in what would be the 'time direction'. As observed in ref. [6], each of the dimensionless coefficients $\alpha_d$ can consistently take only a discrete set of values as a result of the requirement that the action must be independent of the way in which the manifolds $\Omega^{d+1}$ are extended into manifolds $M^{d+2}$ included into $M^{D+1}$ such that $\partial \Omega^{d+1} \equiv M^{d+2}$.[5]

It seem to us that even though any symmetric trace is allowed at this level, if we are looking for a effective theory corresponding to some underlying pregeometric theory then Chern characters may be favored on the grounds of their additivity under a Stieffel (direct) sum of bundles, which would correspond to overlapping 'brane elements' at a point. However we will keep our considerations as general as possible in what follows.

Suppose now that we allow for the possibility that the extended objects with worldvolume $\Omega^{d+1}$ have boundaries $S^d \equiv \partial \Omega^{d+1}$. In that case the action is not gauge invariant anymore, because of a boundary term in the gauge variation $\delta_G$ of the Chern-Simons actions [16][17]. Locally

$$\delta_G I_{d+1}^0(F,A) = dI_d^1(F,A,\lambda) \ , \tag{4}$$

where we are using the notation conventions of ref. [14]. One way to cancel this variation (actually also the global variation) is to add to the action boundary terms of a form analogous to the one in [14][6] (see also ref. [18]) so that the complete action now reads

$$S = \sum_{d \text{ even}}^{D-1} \alpha_d \bigg\{ \int_{\Omega^{d+1}} I_{d+1}^0 + \left[ \tfrac{1}{2} \int_{S^d} d^d \zeta_{(d)} \sqrt{-\gamma_{(d)}} \left( \gamma_{(d)}^{ij} \operatorname{STr}(J_i J_j) - (d-2) \right) \right.$$

---

[5] For instance $P(F) = \operatorname{STr}[e^{iF/(2\pi)}]$ would work.

[6] It is important however to point out that our action is not just the action of [14] for a supergroup, as for our action there is no background metric or Ramond-Ramond (RR) p-form fields, and furthermore our branes do have a bulk with non trivial degrees of freedom and a nontrivial dynamics while the kinetic term 'lives' at the boundary. The branes of [14] would correspond to the boundary of ours in the pure gauge case (when bulk degrees of freedom go away), and if we use the WZNW term from $A$ to mimic the RR field.

$$-\int_{S^d}(C_d(A,K)-b_d)\Big]\Big\} \ , \tag{5}$$

where the $\zeta$'s are local coordinates in $S^d$ and

$$J_i^I = A_m^I \frac{\partial X^m}{\partial \zeta^i} - K_\iota^I \frac{\partial y^\iota}{\partial \zeta^i} \ , \quad \text{(or } J = A - K \text{ as 1-form)} \tag{6}$$

where $K$ is the left-invariant Maurer-Cartan form in the (super)group manifold G, the $y^\iota$'s are coordinates on that (super)group manifold, while $K_\iota(y) = g^{-1}(y)\partial g(y)/\partial y^\iota$ satisfies $dK + K^2 = 0$ (the differentials are taken in $S^d$ for $y$-dependent quantities), and $b_d$ is a $d$-form such that $db_d = -I_{d+1}^0(0,K)$, while $C_d(A,K)$ is chosen in order to preserve gauge invariance in a way described below. The auxiliary metrics $\gamma_{(d)}$ have purely algebraic equation of motion that can be used to eliminate it from the action. The $b_d$-term is a generalization of Wess-Zumino-Novikov-Witten (WZNW)-term. For example, the $d=2$ case corresponds to the usual (super)string formulation [15] on $d=2$ world-sheet with an additional 3D bulk term with $I_3^0$. We must add to the list of our dynamical variables the functions $X_{(d)}^m(\zeta_{(d)}^i)$ (or equivalently $\chi_{(d)}^k(\zeta_{(d)}^i)$), $y_{(d)}^\iota(\zeta_{(d)}^i)$ and $\gamma_{(d)}^{ij}(\zeta_{(d)}^k)$.

Under (super)gauge transformations, each term of our action transforms as [14]

$$\delta_\mathrm{G} A = d\lambda + [A,\lambda] \ , \quad \delta K = d\lambda + [K,\lambda] \ , \tag{7}$$

and accordingly $J$ transforms covariantly

$$\delta_\mathrm{G} J = [J,\lambda] \ . \tag{8}$$

Also

$$\delta_\mathrm{G} I_{d+1}^0(F,A) = dI_d^1(F,A,\lambda) \ , \quad \delta_\mathrm{G} b_d = -I_d^1(0,K,\lambda) \ , \tag{9}$$

and we require

$$\delta_\mathrm{G} C_d = I_d^1(F,A,\lambda) - I_d^1(0,K,\lambda) + \text{(exact form)} \ , \tag{10}$$

which is satisfied for [14]

$$C_d = k_{01} I_{d+1}^0(F_t,A_t) = \int_0^1 dt\, l_t\, I_{d+1}^0(F_t,A_t) \ , \tag{11}$$

where we identify

$$A \equiv A_1 \ , \quad K \equiv A_0 \ , \tag{12}$$

for

$$A_t = tA_1 + (1-t)A_0 \ , \quad F_t = dA_t + A_t^2 = tF + t(t-1)(A_1 - A_0)^2 \ . \tag{13}$$

For $d=2$ for instance, $C_2 = \mathrm{STr}\,(AK)$. The $l_t$ in eq. (11) is defined to act on arbitrary polynomials

$$l_t A_t = 0 \ , \quad l_t F_t = A_1 - A_0 \ , \tag{14}$$

with the convention that $l_t$ is defined to act as an antiderivation. Then the Cartan homotopy operator is defined by [17]

$$k_{01}\mathcal{P}(F_t, A_t) = \int_0^1 \delta t \; l_t \mathcal{P}(F_t, A_t) \;, \tag{15}$$

where $\delta t$ is for the $t$-integration, to be distinguished from other one-forms such as $dA_t$ [17]. By integrating

$$(l_t d + d l_t)\mathcal{P}(F_t, A_t) = \tfrac{\partial}{\partial t}\mathcal{P}(F_t, A_t) \tag{16}$$

over $t$ from 0 to 1, we get the Cartan homotopy formula [17]

$$(k_{01}d + dk_{01})\mathcal{P}(F_t, A_t) = \mathcal{P}(F_1, A_1) - \mathcal{P}(F_0, A_0) \;. \tag{17}$$

Putting all this together we see that the action (5) is indeed gauge invariant.

A more compact form of the action (5) can be given by considering the Cartan homotopy formula for $\mathcal{P}(F_t, A_t) = I^0_{d+1}(F_t, A_t)$, then the action (5) is rewritten as

$$S = \sum_{\substack{d=0 \\ d \text{ even}}}^{D-1} \alpha_d \left\{ \int_{\Omega^{d+1}} k_{01} \operatorname{STr}\left(F_t^{\frac{d+2}{2}}\right) + \tfrac{1}{2} \int_{S^d} d^d \zeta_{(d)} \sqrt{-\gamma_{(d)}} \left[ \gamma^{ij}_{(d)} \operatorname{STr}(J_i J_j) - (d-2) \right] \right\} \;. \tag{18}$$

Here we used also the relationship $\int_{\Omega^{d+1}} I^0_{d+1}(0, K) = -\int_{S^d} b_d$. This form of the action makes its gauge invariance more manifest, because $k_{01}$ is a gauge invariant operator. We can generalize our action one step further, and treat $A_0 = K$ and $A_1 = A$ in a symmetric fashion by letting $A_0$ be an arbitrary gauge field and not just pure-gauge, then the action (5) is just a particular case of the action (18). The price we have paid is doubling the number of bulk degrees of freedom.[7]

It is worthwhile to note that the $\operatorname{STr}(J^2)$ in the 'kinetic' term in the action (5) is gauge invariant by itself, while the bulk and the other boundary terms transform into each other under gauge transformations. That would in principle mean that the coefficient of that term can be chosen independently. However, as will be seen in the next section, the $\kappa$-symmetric Green-Schwarz string action obtained as a certain limit fixes that coefficient, because the $\kappa$-symmetry mixes the kinetic and WZNW-term of the action. In the standard formulation of the heterotic string, we have the choice between a 'fermionic' [19] and a 'bosonic' [20] formulation, depending on whether the gauge group acts on world-sheet fermions or on bosonic group manifold coordinates. Our action is analogous to the latter, but we might think of an alternative fermionic formulation in which the quantum anomaly of a quadratic kinetic boundary term cancels out the boundary gauge variation of the Chern-Simons bulk term, and the WZNW-term is a quantum effect, so that all relative coefficients are determined.

---

[7]The Cauchy problem (as well as causality issues, considering the absence of a background metric) for the action (18) seems formidable, though probably not much worse than the same problems for standard Chern-Simons supergravities. In the case that the manifold in which all the branes are embedded has boundaries, we would need boundary conditions for the gauge fields, as well as specifying whether the extended objects have boundaries on that boundary or not, in order to have a well posed Cauchy problem.

The $\kappa$-simmetry should in principle rule out the infinite number of generally covariant and gauge invariant terms that we can build out of the gauge covariant objects $F$ and $J$ (and their pull-backs) and the gauge invariant $\gamma$'s. However, some alternative forms of the kinetic term may also yield $\kappa$-symmetric actions, for instance Born-Infeld-like kinetic terms as

$$\int_{S^d} d^d \zeta_{(d)} \operatorname{STr} \left[ \sqrt{-\operatorname{sdet} \{J_i J_j + (F_0)_{ij} + (F_1)_{ij}\}} \right] \tag{19a}$$

or

$$\int_{S^d} d^d \zeta_{(d)} \operatorname{STr} \left[ \sqrt{-\operatorname{sdet} \{\operatorname{STr}(J_i J_j) + (F_0)_{ij} + (F_1)_{ij}\}} \right] \tag{19b}$$

where the superdeterminant is taken in the curved indices $i,j$ of the pull-backs on $S^d$ while the supertraces are taken on the group indices. Notice that because supersymmetry is realized as a part of the gauge group for our models the kinetic terms given before are at once a gauge non-abelian and fermionic supersymmetric (as well as generally covariant) generalization of the bosonic abelian Born-Infeld action. Out of these possible alternative kinetic terms, the second one is the one that corresponds to a more straightforward generalization of the metric-eliminated version of our original one. The above actions with Born-Infeld-like kinetic terms would provide us a framework where fundamental strings and Dp-branes can be treated in an equivalent fashion. Duality arguments would likely help in fixing the precise form of the kinetic term. For a review of the literature on D-branes and Born-Infeld terms see [21] and references therein. (Cf. A related recent work in [22])

Notice that our actions of eqs. (5) and (18) describe a system of any number of interacting branes with or without boundaries which are charged with respect to the gauge fields of the model. The currents carried by the branes can be computed by taking the functional derivatives of different pieces of the action with respect to the gauge potentials.

## 3. From $OSp(32|1)$ to Type IIA Green-Schwarz Superstring

We next give some intermediate steps to obtain the action for 10D Type IIA Green-Schwarz superstring from a Maurer-Cartan form for the M-theory group $OSp(32|1)$ in 11D [12][23][9]. Subsequently, we consider how we can understand the usual fermionic $\kappa$-symmetry [24] for the Green-Schwarz superstring [15] in our formulation. We hope this provides us with a clearer picture about the link between our extended objects descrbed by our action (5) or (18) on Chern-Simons supergravity [4][5][6][9] with 10D Green-Schwarz superstring [15].

An earlier work towards establishing a link between Chern-Simons theory for a super-group and superstring theory has been done by Green in [7] (Cf. also [8]). However, several differences with our model should be clarified here:
(i) In the last section, a Chern-Simons action for a supergroup extension of the super Poincaré group is discussed for a three dimensional base manifold, however the action does not contain boundary terms which would render it gauge invariant without gauge-fixing in the boundary. Also a gauge-invariant kinetic term is absent in [7].
(ii) In this section, the WZNW action corresponding to the Chern-Simons action of Section

2 will be considered with a kinetic term now included, but in such a way that the model is again not gauge invariant without gauge-fixing.

(iii) Our branes are embedded in a larger spacetime and interacting with other branes, nemely, the pure gauge condition is not required for us in general, as it is for ref. [7]. That means that we can consider generic backgrounds, and not only flat superspace.

Consider the group $G$ as the M-theory group $OSp(32|1)$ with generators $P_a$, $Q_\alpha$, $M_{ab}$ and $Z_{a_1...a_5}$, and the symmetric trace is just the standard symmetricized supertrace in the adjoint representation of $G$. If we consider the case of 'elementary' fundamental objects of size one in proper coordinates (as opposite to 'cosmic' extended objects), living in a large background[8] then dimensional arguments show that the $n$-dimensional integrals in our action scale as $\mathcal{O}(1/L^n)$ where $L$ is the characteristic scale of the background. That means that the dominant terms in the action will be those of lower dimensions that are non-vanishing. In the case of the group $OSp(32|1)$ the generators are traceless in the adjoint representation, so that the 0-brane (particle) terms are absent and the dominant contribution comes from the 2-brane. We notice that if $A = 0$, and the pure-gauge $K$ is of the form $K = g^{-1}dg$, then we compute essentially up to $\mathcal{O}(R^{-1/2})$-terms. This is equivalent to keeping only $(X^a, \theta^\alpha)$ among the coordinates $(y^\iota)$. Plugging this ansatz back into the action, restricting the boundary of the 2-brane to a 10D boundary of the 11D base manifold, identifying the gauge parameters $X^a$ with the $D = 11$ coordinates, we get the kinetic term from $\text{STr}(J^2)$ in the original action, and the WZNW-term from the $b_2$-term.

The $OSp(32|1)$ algebra is dictated by [25][26]

$$\{Q_\alpha, Q_\beta\} = a_2 \frac{1}{R}(\widehat{\gamma}^{\hat{a}\hat{b}})_{\alpha\beta} \widehat{M}_{\hat{a}\hat{b}} + ia_5 \frac{1}{\sqrt{R}}(\widehat{\gamma}^{\hat{a}_1\cdots\hat{a}_5})_{\alpha\beta} \widehat{Z}_{\hat{a}_1\cdots\hat{a}_5} \ , \tag{20a}$$

$$[\widehat{M}_{\hat{a}\hat{b}}, Q_\alpha] = -\tfrac{1}{2}(\widehat{\gamma}_{\hat{a}\hat{b}})_\alpha{}^\beta Q_\beta \ , \qquad [\widehat{M}_{\hat{a}\hat{b}}, \widehat{M}^{\hat{c}\hat{d}}] = -4\delta_{[\hat{a}}{}^{[\hat{c}} \widehat{M}_{\hat{b}]}{}^{\hat{d}]} \ , \tag{20b}$$

$$[\widehat{Z}_{\hat{a}_1\cdots\hat{a}_5}, Q_\alpha] = ib_3 \frac{1}{\sqrt{R}}(\widehat{\gamma}_{\hat{a}_1\cdots\hat{a}_5})_\alpha{}^\beta Q_\beta \ , \qquad [\widehat{M}_{\hat{a}\hat{b}}, \widehat{Z}^{\hat{c}_1\cdots\hat{c}_5}] = -10\delta_{[\hat{a}}{}^{[\hat{c}_1} \widehat{Z}_{\hat{b}]}{}^{\hat{c}_1\cdots\hat{c}_5]} \ , \tag{20c}$$

$$[\widehat{Z}_{\hat{a}_1\cdots\hat{a}_5}, \widehat{Z}_{\hat{b}_1\cdots\hat{b}_5}] = \frac{1}{\sqrt{R}} f_{\hat{a}_1\cdots\hat{a}_5\ \hat{b}_1\cdots\hat{b}_5}{}^{\hat{c}_1\cdots\hat{c}_5} \widehat{Z}_{\hat{c}_1\cdots\hat{c}_5} \ , \tag{20d}$$

in the $SO(1,10)$ manifest notation. The $a$'s and $b$'s are appropriate constants, and $f$'s in (20d) is a structure constant for the $[\widehat{Z}, \widehat{Z}]$-commutator, whose detail is not important here. Here all the indices and operators with *hats* refer to the 11D,[9] to be distinguished from those in 10D into which we are now going to perform the compactification on a torus from the former. The constant $R$ denotes the radius of such a torus. We assigned appropriate powers of $R$ in such a way that in the limit $R \to \infty$ recovers the usual super Poincaré algebra in 10D, with the consistent mass dimensions, namely $[\widehat{P}_{\hat{a}}]_M = +1$, $[\widehat{M}_{\hat{a}\hat{b}}]_M = 0$, $[\widehat{Z}_{\hat{a}_1\cdots\hat{a}_5}]_M = +1/2$, $[Q_\alpha]_M = +1/2$. For example, the half-integer for $[\widehat{Z}]_M = +1/2$ is required by the vanishing of the $\widehat{Z}$-term in (20a) for $R \to \infty$.

Upon the compactification for a finite $R$, the algebra above becomes what is called 10D de-Sitter algebra [26], where in particular, the components of $\widehat{M}_{\hat{a}\hat{b}}$ with the 10-th coordinate,

---

[8]Notice that this is *a priori* a somewhat unnatural limit in our approach.

[9]For $Q$'s we do not need this distinction, because all of its Majorana components are kept in 10D upon the compactification. Relevantly, the indices $\alpha, \beta, \cdots$ are Majorana indices in 10D.

are identified with the momentum operator $P_a$ in 10D:

$$\widehat{M}_{\hat{a}\hat{b}} = \begin{cases} \widehat{M}_{ab} = M_{ab} \;, \\ \widehat{M}_{10b} = RP_b = -\widehat{M}_{b\,10} \;. \end{cases} \tag{21}$$

Thus the algebra (20) becomes the de-Sitter algebra in 10D [26]:

$$\{Q_\alpha, Q_\beta\} = ia_1(\gamma_{11}\gamma^c)_{\alpha\beta}P_c + a_2\tfrac{1}{R}(\gamma^{ab})_{\alpha\beta}M_{ab} + ia_5\tfrac{1}{\sqrt{R}}(\gamma^{a_1\cdots a_5})_{\alpha\beta}Z_{a_1\cdots a_5} \;, \tag{22a}$$

$$[M_{ab}, M^{cd}] = -4\delta_{[a}{}^{[c}M_{b]}{}^{d]} \;, \qquad [M_{ab}, P_c] = +2\eta_{c[b}P_{a]} \;, \tag{22b}$$

$$[P_a, P_b] = ib_1\tfrac{1}{R^2}M_{ab} \;, \quad [M_{ab}, Q_\alpha] = -\tfrac{1}{2}(\gamma_{ab})_\alpha{}^\beta Q_\beta \;, \quad [P_a, Q_\alpha] = ib_2\tfrac{1}{R}(\gamma_{11}\gamma_a)_\alpha{}^\beta Q_\beta \;, \tag{22c}$$

$$[\widehat{Z}_{\hat{a}_1\cdots\hat{a}_5}, Q_\alpha] = ib_3\tfrac{1}{\sqrt{R}}(\widehat{\gamma}_{\hat{a}_1\cdots\hat{a}_5})_\alpha{}^\beta Q_\beta \;, \quad [\widehat{M}_{\hat{a}\hat{b}}, \widehat{Z}^{\hat{c}_1\cdots\hat{c}_5}] = -10\delta_{[\hat{a}}{}^{[\hat{c}_1}\widehat{Z}_{\hat{b}]}{}^{\hat{c}_2\cdots\hat{c}_5]} \;, \tag{22d}$$

$$[P_b, \widehat{Z}^{\hat{a}_1\cdots\hat{a}_5}] = -5\tfrac{1}{R}\delta_{10}{}^{[\hat{a}_1}\widehat{Z}_b{}^{\hat{a}_2\cdots\hat{a}_5]} \;, \quad [\widehat{Z}_{[\hat{5}]}, \widehat{Z}_{[\hat{5}]'}] = \tfrac{1}{\sqrt{R}}f_{[\hat{5}][\hat{5}]'}{}^{[\hat{5}]''}\widehat{Z}_{[\hat{5}]''} \;, \tag{22e}$$

where all other independent commutators are vanishing. The index $[\hat{5}]$ on $Z_{[\hat{5}]}$ etc. is for the antisymmetric indices $\hat{a}_1\hat{a}_2\cdots\hat{a}_5$. We did not decompose the indices in terms of 10D indices for (22e), because they are not crucial for our purpose of getting the expression for $g^{-1}dg$ up to $\mathcal{O}(R^{-3/2})$. The reason we need an extra $\gamma_{11}$ in (22a) is explained in [27].

In the limit $R \to \infty$ the algebra (22) recovers the usual super Poincaré algebra in Minkowski 10D, so that we can see the link with the Green-Schwarz superstring action [15].

We now compute the Maurer-Cartan form $K = g^{-1}dg$ up to $\mathcal{O}(R^{-3/2})$, following the general expansion formula:

$$K = g^{-1}dg = \sum_{n=1}^{\infty} \tfrac{1}{n!} \overbrace{[[\cdots[[}^{n} d, \overbrace{h], h], \cdots, h], h]}^{n} \;, \tag{23}$$

where $g = \exp(iX^a P_a + \theta^\alpha Q_\alpha) \equiv \exp h$. We keep at most the $\mathcal{O}(R^{-1/2})$ and $\mathcal{O}(R^{-1})$-terms, in order to elucidate the first effect of our 11D model on the Green-Schwarz superstring action. As a simple dimensional consideration, as well as the commutators that are vanishing, we see that all the terms at $\mathcal{O}(R^{-1})$ are exhausted at the fourth commutator $(n = 4)$ in (22).

Considering all of these, we can get the expression

$$\begin{aligned} g^{-1}dg =\; & +idX^a P_a + d\theta^\alpha Q_\alpha \\ & + \tfrac{1}{2}\Big[ +ia_1(d\bar{\theta}\gamma^a\theta)P_a + ia_5\tfrac{1}{\sqrt{R}}(d\bar{\theta}\gamma^{[5]}\theta)Z_{[5]} - b_2\tfrac{1}{R}dX^a(\bar{\theta}\gamma_a)^\alpha Q_\alpha \\ & \qquad + b_2\tfrac{1}{R}(d\bar{\theta}\gamma_a)^\alpha X^a Q_\alpha + a_2\tfrac{1}{R}(d\bar{\theta}\gamma^{cd}\theta)M_{cd} \Big] \\ & + \tfrac{1}{6}\Big[ -b_2\tfrac{1}{R}(d\bar{\theta}\gamma^b\theta)(\bar{\theta}\gamma_b)^\alpha Q_\alpha - ia_1b_2\tfrac{1}{R}(d\bar{\theta}\gamma_b\gamma^a\theta)X^b P_a + ia_5b_3\tfrac{1}{R}(d\bar{\theta}\gamma^{[5]}\theta)(\bar{\theta}\gamma_{[5]})^\alpha Q_\alpha \\ & \qquad + ia_1b_2\tfrac{1}{R}dX^a(\bar{\theta}\theta)P_a - 2ia_2\tfrac{1}{R}(d\bar{\theta}\gamma^{ab}\theta)X_a P_b - i\tfrac{1}{2}a_2\tfrac{1}{R}(d\bar{\theta}\gamma^{bc}\theta)(\bar{\theta}\gamma_{bc})^\alpha Q_\alpha \Big] \\ & + \tfrac{1}{24}\Big[ -ib_2a_1\tfrac{1}{R}(d\bar{\theta}\gamma^a\theta)(\bar{\theta}\theta)P_a + \tfrac{1}{2}a_1a_2\tfrac{1}{R}(d\bar{\theta}\gamma^{bc}\theta)(\bar{\theta}\gamma_{bc}\gamma^d\theta)P_d \\ & \qquad + a_1a_5b_3\tfrac{1}{R}(d\bar{\theta}\gamma^{[5]}\theta)(\bar{\theta}\gamma_{11}\gamma_{[5]}\gamma^a\theta)P_a \Big] + \mathcal{O}(R^{-3/2}) \;. \end{aligned} \tag{24}$$

All of the $\mathcal{O}(R^{-1/2})$ or higher-order terms will disappear in the limit $R \to \infty$, and we are left with the standard factor in the Green-Schwarz $\sigma$-model formulation:

$$K \approx i(dX^a + \tfrac{1}{2}a_1 \bar\theta \gamma^a d\theta)P_a + d\theta^\alpha Q_\alpha \ . \tag{25}$$

For deriving the Type IIA Green-Schwarz action from ours, the relevant traces of products of generators in the adjoint representation of $OSp(32|1)$ are $\mathrm{STr}\,(PP)$, $\mathrm{STr}\,(PQ)$ and $\mathrm{STr}\,(QQ)$, while for the WZNW-term the relevant traces are $\mathrm{STr}\,(PPP)$, $\mathrm{STr}\,(PPQ)$, $\mathrm{STr}\,(QQP)$ and $\mathrm{STr}\,(QQQ)$. Among these, the non-vanishing ones are normalized as

$$\mathrm{STr}\,(P_a P_b) = \eta_{ab} \ , \qquad \mathrm{STr}\,(P_a Q_\alpha Q_\beta) = \tfrac{i}{2}a_1 (\gamma_{11}\gamma_a)_{\alpha\beta} \ . \tag{26}$$

In the process of getting the action for 10D Type IIA Green-Schwarz superstring [15], the usual fermionic $\kappa$-symmetry [24], in particular how to understand it in our formulation, plays a role of an important guiding principle. We regard the $\kappa$-symmetry as a restricted gauge transformation given like (7) or (8), under the condition that the gauge field is vanishing: $A = 0$, which is stronger than a pure-gauge condition, while acting only on $K$, but not on $A$, as opposed to eq. (7). Since $K$ is a pure-gauge, its form is to be the same as (24), after the limit $R \to \infty$. We can identify this $K$ with the 'pull-back' used in the Green-Schwarz $\sigma$-model [15]: $K_i{}^A = \Pi_i{}^A \equiv (\partial_i Z^M) E_M{}^A$, where we identify the index index $\alpha$ for $Q_\alpha$ with that of the fermionic coordinates in superspace. In fact, as a simple algebra reveals, the pull-back $\Pi_i{}^A$ satisfies the pure-gauge condition $d\Pi + \Pi^2 = 0$ on a flat 10D background, so that this identification is consistent.

Thus the $\kappa$-symmetry acts on $K$ similarly to (7), but leaving $A$ intact:

$$\delta_\kappa K = d(\delta_\kappa h) + [K, \delta_\kappa h] \ , \qquad \delta_\kappa A = 0 \ , \tag{27}$$

where $\delta_\kappa h \equiv g^{-1}\delta_\kappa g$. This also maintains the pure-gauge condition of $K$. To be more specific, we have

$$\delta_\kappa h = g^{-1}\delta_\kappa g = [\,(\delta_\kappa X^m)E_m{}^a + (\delta_\kappa \theta^\mu)E_\mu{}^a\,]\,P_a + (\delta_\kappa \theta^\mu)E_\mu{}^\alpha Q_\alpha \ . \tag{28}$$

As usual, the $P_a$-term vanishes under the condition $\delta_\kappa X^m = \tfrac{i}{2}a_1[(\delta_\kappa \bar\theta)\gamma_{11}\gamma^m \theta]$, while the $Q_\alpha$-term corresponds to the usual $\kappa$-parameter, as a shift of the fermionic coordinates $\theta^\mu$. In this sense, $\delta_\kappa h$ has only the fermionic component: $\delta_\kappa h = (\delta_\kappa h)^\alpha Q_\alpha$.

In order to get an explicit link with the Green-Schwarz superstring action, we take the limit $R \to \infty$, and restrict the original $OSp(32|1)$ group down to the super Poincaré group. This corresponds to a truncation of components in the $\mathrm{STr}$-operation. We use the symbol $\mathrm{STr}'$ for such a truncated trace, in which the generators in fermionic contraction $K_i^\alpha K_{j\alpha}$ are excluded. By this restriction, the original $OSp(32|1)$ gauge invariance is lost, but the action $S_{\mathrm{GS}}$ has the fermionic $\kappa$-symmetry instead as a reminiscent of the original $OSp(32|1)$ gauge invariance. Applying this prescription to our action S in (18) for $d = 2$ (with $\alpha_d = 0$ for $d \neq 2$), we get

$$S_{\mathrm{GS}} = \int_{S^2} \left[\, d^2\zeta\, \tfrac{1}{2}\sqrt{-\gamma}\gamma^{ij}\,\mathrm{STr}'(K_i K_j) + b_2 \,\right] = \int_{S^2} \left[\, d^2\zeta\, \tfrac{1}{2}\sqrt{-\gamma}\gamma^{ij} K_i^a K_{ja} + b_2 \,\right] \ . \tag{29}$$

We can replace this sort of rather 'artificial' restriction of STr by considering some rescaling of the algebra (23) after the $R \to \infty$. To be specific, if we rescale $P_a$ and $Q_\alpha$ by

$$P_a = \xi P'_a \ , \qquad Q_\alpha = \sqrt{\xi} Q'_\alpha \ , \tag{30}$$

putting also an additional rescaling factor $\xi^{-2}$ in front of our action (5), and take the limit $\xi \to \infty$, then we get exactly the same action as (29): $\lim_{\xi \to \infty} \lim_{R \to \infty} S = S_{\mathrm{GS}}$ for $d = 2$. Namely, the restricted supertrace $\mathrm{STr}'$ can be understood as such a limiting process. Of course we can just *define* the supergroup $G$ to be super Poincaré and the invariant tensor in the superalgebra to be given by the trace just considered. At any rate, our point in obtaining it from the $OSp(32|1)$ model is to show that it appears as the result of a limiting process, not that the former is equivalent to the later.

The $\kappa$-invariance of $S_{\mathrm{GS}}$ can be investigated more explicitly: First, we note the gauge transformation of $b_2$ is

$$\delta_{\mathrm{G}} b_2 = -I_2^1(0, K, \lambda) = +\mathrm{STr}\,(K^2 \lambda) \ . \tag{31}$$

Second, since our $\kappa$-transformation is a restricted gauge transformation of $K$, we can read from (31) that

$$\begin{aligned}
\delta_\kappa b_2 &= d\zeta^i \wedge d\zeta^j \, \mathrm{STr}\,\Big( K_i^a P_a [\, K_j^\alpha Q_\alpha \, , \, (\delta_\kappa h)^\beta Q_\beta \,] \Big) \\
&= -i a_1 \epsilon^{ij} d^2\zeta \, K_i^a K_j^\alpha (\gamma_{11} \gamma_a)_{\alpha\beta} (\delta_\kappa h)^\beta \ ,
\end{aligned} \tag{32}$$

and similarly,

$$\begin{aligned}
\delta_\kappa \left[ \tfrac{1}{2} \sqrt{-\gamma} \gamma^{ij} K_i^a K_{ja} \right] &= \sqrt{-\gamma} \gamma^{ij} K_i^a \Big[ i a_1 K_j^\alpha (\gamma_{11} \gamma_a)_{\alpha\beta} (\delta_\kappa h)^\beta \Big] + \tfrac{1}{2} [\delta_\kappa(\sqrt{-\gamma} \gamma^{ij})] \, K_i^a K_{ja} \\
&= i a_1 \sqrt{-\gamma} \gamma^{ij} K_i^a K_j^\alpha (\gamma_{11} \gamma_a)_{\alpha\beta} (\delta_\kappa h)^\beta + \tfrac{1}{2} [\delta_\kappa(\sqrt{-\gamma} \gamma^{ij})] \, K_i^a K_{ja} \ , \quad (33)
\end{aligned}$$

where we have used (26). As is easily seen now, the addition of (32) to (33) yields $\delta_k S_{\mathrm{GS}} = 0$. In fact, if we set up the $\kappa$-transformation [28] as

$$\delta_\kappa h = i \Pi_-{}^a (\gamma_a \kappa_+)^\alpha Q_\alpha \equiv i(\slashed{\Pi}_- \kappa_+)^\alpha Q_\alpha \ ,$$
$$\delta_\kappa V_+{}^i = 4 a_1 \Pi_+{}^\alpha (\gamma_{11})_\alpha{}^\beta \kappa_{+\beta} \equiv 4 a_1 (\overline{\Pi}_+ \gamma_{11} \kappa_+) \ , \qquad \delta_\kappa V_-{}^i = 0 \ , \qquad \delta_\kappa \sqrt{-\gamma} = \delta_\kappa V = 0 \ , \tag{34}$$

where $+, -$ are local Lorentz indices in the light-cone frame, $V_\pm{}^i$ are the zweibeins: $\gamma^{ij} \equiv V_+{}^j V_-{}^j + V_-{}^i V_+{}^j$, and we identified $K_i^A \equiv \Pi_i{}^A$. Now we easily see that

$$\begin{aligned}
\delta_\kappa \left[ \tfrac{1}{2} \sqrt{-\gamma} \gamma^{ij} \Pi_i{}^a \Pi_{ja} \right] &= -2 a_1 V (\overline{\Pi}_+ \gamma_{11} \kappa_+) \Pi_-{}^a \Pi_{-a} \ , \\
\delta_\kappa b_2 &= +2 a_1 V (\overline{\Pi}_+ \gamma_{11} \kappa_+) \Pi_-{}^a \Pi_{-a} \ ,
\end{aligned} \tag{35}$$

*i.e.*, $\delta_\kappa S_{\mathrm{GS}} = 0$ for the total action in the Type IIA Green-Schwarz superstring [15][28]. Notice that the peculiar $\mathrm{STr}'$-operation excludes the derivative term $d(\delta_\kappa h)^\alpha Q_\alpha$ as usual in the Green-Schwarz formulation [15].

## 4. Concluding Remarks

In this Letter, we have proposed a new action for extended objects on Chern-Simons supergravity [4][5][6][9], and discussed its possible link with 10D Type IIA Green-Schwarz superstring theory [15]. Our action describe in general a system of interacting branes with or without boundaries with a gauge symmetry of supersymmetric type and generally covariant. The equations of motion of the system can be interpreted as if the different branes carry distributional charges in their world-volumes.

As for the quantization of this system, we can quantize it formally in the path integral approach where we should in principle sum over all possible configurations and topologies of this system. However such an ungrateful task might not be required, as we may expect on the basis of the (presumed) absence of counterterms[10] with the required symmetries that our action is in fact the effective action already, then all what we need is to consider a background field perturbative expansion around a solution of the classical equations of motion (the vacuum).[11]

We may also get the link with Type IIB Green-Schwarz superstring [15]. To this end, we have to consider a 11D slab with two 10D boundaries for which we identify the gauge parameters with the 10D coordinates. We chose one half of the 32 components of the spinor to be zero by imposing a Weyl condion with respect to the $\gamma_{11}$ so the spinor is a 10D Majorana-Weyl spinor. Both 10D boundaries are going to be 'identified' in the sense that we are going to assume that the gauge parameters $X^a$ correspond to the same 10D space, however that identification can be done in two ways, related by parity. Those ways are equivalent regarding the $X^a$, but the spinors in each boundary will have different chiralities. To be more specific, suppose we have two parallel boundaries. The 11D Majorana spinor has 32 components on the first boundary with the components splitting as $(\mathbf{32}) = (\mathbf{16}_L, \mathbf{16}_R)$, where $_L$ and $_R$ denote the chiralities in 10D. Similarly on the second boundary, a Majorana spinor has $(\mathbf{32'}) = (\mathbf{16'}_L, \mathbf{16'}_R)$. Among these four kinds of Majorana-Weyl spinors on the two 10D boundaries 'identified', we can truncate either the combination $\mathbf{16}_R, \mathbf{16'}_L$ or $\mathbf{16}_R, \mathbf{16'}_R$. We have in each case Type IIA and Type IIB superstring actions, respectively.

Notice that even though the reduced action looks like an action of a string embedded into flat superspace, our conceptual frame is quite different, because the Green-Schwarz coordinates are gauge parameters in our case.

We believe that heterotic strings could be also obtained, if we take the $y$'s to be nonzero and combine $\theta$'s and $y$'s to get commuting objects to pass from $Sp(32)$ to $SO(32)$, and alternatively from $Sp(16) \times Sp(16)$ to $SO(16) \times SO(16)$ which combined with a similar 'bosonization' of halves of $Q_\alpha$ gives $E_8 \times E_8$. Of course we could get the gauge groups just by adding them to the $OSp(32|1)$ supergroup, because we have no constraints so far on which group we use, but that would not be so terse conceptually. Type I strings do presumably corresponds to the the degenerate case of a flattened membrane.

---

[10]Possibly assuming also additivity under Stieffel sum and the above mentioned quantization of the $\alpha$'s.

[11]A candidate to a vacuum that seems interesting is $AdS^4 \times S^7$ with the spatial sections of the branes contained into the $S^7$ part, except maybe for the 8-brane which may be absent or have four of its spatial dimensions in each sector. It may be that the interaction between the branes is responsable for the smallness of the $S^7$ part.

We could also consider adding a non pure-gauge contribution to the potential and dimensionally reduce it by assuming the gauge potential is independent of the membrane and background coordinates across the slab. Then proceeding in the same way as in ref. [4] in reducing the 3D Chern-Simons action to a 2D topological action, we would have (for some choices of our symmetric supertrace only) an additional term in the string action that looks like the dilaton times the curvature scalar of the string world-sheet, where the dilaton is essentially the eleventh component of the vielbein and the 2D curvature tensor is the one obtained pull-back of the 11D one. This seems to provide a heuristic link to the relationship 11D size-dilaton-coupling strength. Notice that even though we can order the contributions of different string configurations to the path integral by the power of that coupling strength and relate that to the genus of the string world-sheet, so that it looks like we have a first quantized theory, we would actually have a second quantized theory, and our classical strings are already first quantized objects.

Chern-Simons supergravity theories [5][6] have so far developed with no explicit direct link with the conventional supergravity or superstring theories [15], in the sense that the translation operators in the former $P_a$ in supergroups do not really generate 'infinite' dimensional diffeomorphism. In other words, all the generators in supegroups generate only 'internal' symmetries, separated from the space-time symmetries. It has been only in a certain speculative limit that the diffeomorphism is supposed to be generated or identified with the $P_a$ operators in Chern-Simons supergravity [9]. An important step in that direction has been taken recently by M. Bañados [29], who added a small cosmological constant term to the Chern-Simons supergravity action in 5D in order to make contact with standard supergravity in 5D. It seems plausible that the extended objects of our model may provide such an additional cosmological constant without breaking any symmetries. Considering also the speculation from different grounds in refs. [6][9][29], the 11D part of our model might have as a low energy truncation of 11D supergravity. As a supporting evidence for this, we have provided a more solid link with 10D superstring [15], by taking an explicit limit to get the Type IIA Green-Schwarz action in flat superspace, with supersymmetry introduced as a part of a supergroup.

In our formulation, we did not introduce the target supergravity. The reason for this is that within our conceptual framework, a natural extension to a more general background is not a string mapped into curved superspace, but the action given by (5) or (18). Even though this looks like a major drawbacks in our formulation at first sight, we emphasize that this situation is similar to the conventional M-theory formulation, in which the particular large $N$-limit in the 1D matrix model is believed to reproduce the curved supergravity background in 11D [30]. In other words, it is natural to expect that our microscopic model of Chern-Simons supergravity may well reproduce the 'ordinary' curved supergravity background for the target space in a certain limit for superstring theory.

It is important to notice that our model is not and does not in any obvious way reduce to supermembrane theory [31], which has closed (2+1)-dimensional objects with a quadratic kinetic term as well as a WZNW-term in the world-volume. However, quoting from the first reference of [12] referring to M-theory [10][11][12], *"... it is not obvious that the presence of a membrane in D=11 implies the existence of a supermembrane theory, ... but it is important to appreciate that, however things turn out, the major premise of M-theory, for which the circumstantial evidence is now overwhelming, is that there exist* **some** *consistent supersymmetric quantum theory in D=11 containing membranes and 5-branes,*[12] *with D=11 supergravity as its effective field theory."*

Considering all the previous points, it may not seem so implausible to conjecture that a model of the kind presented here for the supergroup $OSp(32|1)$ and 11D may actually correspond to M-theory [10][11][12], understood as the continuum limit of the underlying pregeometric theory we have mentioned before, which could be called 'covariant matrix theory'.

We also stress that our formulation will provide us with a link between Chern-Simons supergravity [5][6][9][29] and the conventional superstring/supergravity or M-theory [10][11][12], in the sense that the Green-Schwarz superstring action for the latter is recovered in an explicit limit from the former, even though for the present time we can show that this process works only for flat backgrounds.

Clearly much more development is expected, especially concerning any constraint in the choice of the supergroup $G$ and the dimensionality of the theory and the extended objects coming from consistency considerations. It might be useful to try to translate the reasoning leading to such constraints in superstring theory into the formulation presented here. It might also be that these questions have a natural answer within the context of a future pregeometric theory having a model of the kind proposed here as its effective theory.

We are grateful to S.J. Gates, Jr., for stimulating discussions, and reading the manuscript. We are also indebted to the referee who helped to improve an earlier version of the paper.

---

[12]The 5-branes are solitonic objects, while the membranes are fundamental objects